\documentclass[twocolumn,prl]{revtex4}

\usepackage[utf8]{inputenc}
\usepackage{amsmath}
\usepackage{amsfonts}
\usepackage[pdftex]{graphicx}
\usepackage{amssymb}
\usepackage[pdftex]{hyperref}
\usepackage{caption}
\usepackage[ruled,linesnumbered]{algorithm2e}
\usepackage{multirow}
\usepackage{parskip}
\usepackage{tabularx}
\usepackage{graphicx,psfrag,fancyhdr,layout,appendix,subfigure}
\usepackage{verbatim}
\usepackage{framed}
\usepackage{amsthm}
\usepackage{booktabs}
\usepackage{placeins}

\makeatother

\newcommand{\R}{\ensuremath{{\sf R\hspace*{-0.9ex}\rule{0.15ex}
{1.5ex}\hspace*{0.9ex}}}}
\newcommand{\cnot}{\leavevmode\hbox{\footnotesize{CNOT }}}
\newcommand{\trace}{\ensuremath{\textsf{Tr}}}
\newcommand{\oth}[3]{\ensuremath{\left( \begin{array}{c} #1 \\ #2 \\ #3
\end{array} \right)}}
\newcommand{\me}{\ensuremath{\mathrm{e}}}
\newcommand{\mi}{\ensuremath{\mathrm{i}}}

\begin{document}

\title{Optimising the Solovay-Kitaev algorithm}
\author{Pham Tien Trung, Rodney Van Meter, Clare Horsman}
\affiliation{Keio University Shonan Fujisawa Campus, 5322 Endo, Fujisawa, Kanagawa, Japan}

\begin{abstract}
The Solovay-Kitaev algorithm is the standard method used for approximating arbitrary single-qubit gates for fault-tolerant quantum computation. In this paper we introduce a technique called \emph{search space expansion}, which  modifies the initial stage of the Solovay-Kitaev algorithm, increasing the length of the possible approximating sequences but without requiring an exhaustive search over all possible sequences. We show that our technique, combined with a GNAT geometric tree search outputs gate sequences that are almost an order of magnitude smaller for the same level of accuracy. This therefore significantly reduces the error correction requirements for quantum algorithms on encoded fault-tolerant hardware.
\end{abstract}

\maketitle

\section{Introduction}

 The biggest challenge to building working quantum computers is the problem of fault-tolerance \cite{PreskillFT}. Unless a quantum computer is specifically designed and built to withstand the effects of errors from environmental decoherence and inaccurate hardware operations, it will not be able to perform computations of any significant size \cite{preskillreliable}. One prominent way to control errors is to correct them when they arise, by encoding quantum data in many physical qubits. If the underlying hardware is accurate to ``threshold" value, then these error correction codes can in principle keep an arbitrary computation error-free given enough physical resources \cite{shor,shordiv,dummies}. 

 One of the drawbacks to error correction codes, however, is that only a very small number of logical gates are available to the code. To implement a given algorithm, the gates in the algorithm must be decomposed into the fundamental ``library" gate set available to the code. In general this cannot be performed exactly -- we must look for sequences of library gates that \emph{approximate} the gates we require \cite{kitaev1}.

The key problem when finding an approximation to an arbitrary gate is the size of the space over which it is necessary to search in order to find a good approximation.  In the general case, the longer the sequence of library gates (and their inverses), the better an approximation to the gate in question can be found. However, high-accuracy exhaustive searches become untenable on current computational technology.  The Solovay-Kitaev approximation algorithm was introduced to get around this difficulty \cite{kitaev1,kitaev2002classical}. This recursive algorithm performs an exhaustive search only over the space of sequences of length up to $l_0$, finding the sequence with the smallest distance (defined by the trace norm) from the gate we wish to approximate. The residual difference between the actual and approximate gate is then sent to the next level of the algorithm, where it is further approximated.  At each level the length of the gate sequences grows by a factor of 5. The algorithm terminates at a ``good" approximation, where the distance to the actual gate is less than a chosen constant $\epsilon$.

While very powerful, the Solovay-Kitaev algorithm suffers from a serious weakness. While it will always find a good approximation for any value of $\epsilon$, the search covers only a very sparse region of the entire space of possible approximation sequences. As a consequence, the output from the approximation is almost always far longer than it needs to be. This is an extreme disadvantage for fault-tolerant computation, as this greatly increases the logical depth of the computation, thus increasing the amount of error correction required, which in turn increases the physical size and run time of the algorithm. Reducing these requirements as far as possible is key to implementing realistic fault-tolerant quantum computation.

Currently there are two alternatives to the original Solovay-Kitaev algorithm. The first is an exhaustive-search algorithm that, in certain cases, can give a very efficient library-gate decomposition \cite{austindecomp}. However, in the general case the scaling is exponential as in a standard exhaustive search and therefore limited by computational ability. The second is the ``phase kickback" method originally given in \cite{kitaev2002classical}. This can produce shorter gate sequences by using special ancilla states, but requires many more qubits for the ancilla and is also restricted in the library gate set it can use.

In this paper we describe an alternative gate decomposition algorithm that modifies the original Solovay-Kitaev method and allows a much denser search of the space of sequences. The key to improving the algorithm is to concentrate on the initial approximation, and we  dramatically increase the search space by combining two sequences out of the database of initial sequences. This expansion technique can be implemented recursively, creating an even better initial approximation. To make this search step computationally tractable, we have applied a geometric nearest-neighbour access tree search procedure (GNAT)\cite{GN:1995}. This procedure is a fully general method for any gate and any library gate set, and with high probability produces approximations that are significantly shorter than those given by the original algorithm, for the same precision. 

The paper is structured as follows. Section 2 reviews the Solovay-Kitaev theorem, along with with the original implementing algorithm proposed in \cite{kitaev1,kitaev2002classical}. In Section 3 we introduce the search space expansion technique (SSE) that we will use to supplement the original algorithm. We show that by splitting each candidate sequence and then searching points near to the subsequences, we can generate candidate sequences that with high probability are a better approximation but which are shorter than those at the next level of recursion. One potential issue with this approach is an added classical search requirement over the sampled sequences, and in Section 4 we describe a way of efficiently searching these sequences to reduce the time required compared with a standard exhaustive search. In Section 5 we compare the original Solovay-Kitaev algorithm with our modified algorithm for a set of 25 single-qubit gate unitaries chosen at random. On average, the length of the candidate sequence reduces by a factor of 7, and for some sequences it can reduce it by an order of magnitude. The number of levels of recursion in the algorithm is also greatly reduced, with generally only two or at most three levels being needed even for very high accuracies. 
\section{The Solovay-Kitaev approximation}

The Solovay-Kitaev theorem tells us we can always approximate a single-qubit gate $G$ to arbitrary accuracy $\epsilon$ with a finite sequence of fundamental gates from a universal ``library set" of gates $\{L_1, L_2, \ldots L_N\}$ and their adjoints. The Solovay-Kitaev algorithm gives a method of finding what these sequences are, for a given $G$, $\epsilon$, and $\{L_i\}$.

The Solovay-Kitaev theorem is invoked as part of the compilation process for fault-tolerant quantum computing. The library set on a fault-tolerant computer is in general very limited, with only a handful of gates able to operate within the code space. For example, in the surface code we have access only to a fundamental library of four gates: $\{\cnot,H,S,T\}$ (where $S$ is the phase gate and $T$ a $Z$-rotation of $\pi/4$)\cite{raussendorfprl}. In general, the library contains a 2-qubit entangling gate such as \cnot or controlled-phase, and then some single-qubit gates. Any multi-qubit gate can be decomposed exactly into a combination of a 2-qubit maximally entangling gate plus arbitrary single qubit rotations  \cite{barenco}, so this becomes the first stage of compilation for an algorithm. The next stage is then the further task of decomposing these arbitrary single qubit gates into the single qubit gates of the error correction code library.

The Solovay-Kitaev theorem states that, for a given gate $G$, accuracy $\epsilon$, and library gate set $\{L_i\}$, there always exists a sequence of library gates $(\Pi_{j=1}^{l} A_j\  | \ A_j \in \{L_i\} \{L^\dagger_i\})$ such that
\begin{equation} || G - \Pi_l A_l || \le \epsilon\end{equation}

\noindent using the standard operator trace norm distance
\begin{equation} || M-N || = \trace\sqrt{(M-N)^\dagger(M-N))}\end{equation}

All gates are represented here by unitary operators in $SU(n)$, where $n$ is the dimensionality of the gate. The theorem further states that the length of the sequences, $l$, varies with the required accuracy $\epsilon$ as $ l = O\left(log^c\left(1/ \epsilon\right)\right)$.

Exactly what the constant c is depends on the particular implementation of the decomposition. It is known that the best possible scaling is $c=1$, but with a non-constructive proof \cite{harrow2002efficient}. The standard algorithm gives a scaling of sequence size with accuracy of $c\approx 4$ \cite{SK:2005}.

The most straightforward procedure for performing such a decomposition is to search over all sequences, beginning with the shortest first, until one is found within the required $\epsilon$ of the gate $G$ being decomposed. Unfortunately, such an exhaustive search becomes untenable very quickly. For a library of $n$ fundamental gates, then the number of sequences of length $l$ comprising these library gates and their adjoints is $2n^l$. For example, if we have a library of 5 fundamental single-qubit gates (as, for example, in the surface code: $H, S,S^\dagger,T,T^\dagger$) then a modern server with 64GB memory could hold only up to sequences of length $l\approx 13$. Searching over this size of database is also a significant classical processing task. The longer the sequences are then the higher the chance of finding a sequence within $\epsilon$ of $G$; without the ability to search longer sequences, the correct accuracy may be unobtainable.

The standard Solovay-Kitaev algorithm uses such an exhaustive search technique at its base layer, but then builds on that recursively. We can describe the algorithm in iterative fashion as follows. The base level approximation comprises a search over the space of all sequences of length up to $l_0$. The closest approximation to $G$ is found, 
\begin{equation} S(0) = \Pi_{j=1}^{l_0} A_j\  | \ A_j \in \{L_i\}\cup \{L^\dagger_i\}\label{so}\end{equation}

\noindent We can then decompose the gate as $ G = U(\delta) S(0)$. The operator $U(\delta)$ is the ``residual" of the approximation: how far away from G the operator sequence $S(0)$ still is.

If $||G-S(0)||\le \epsilon$ then the algorithm terminates here and returns $S(0)$ as the appropriate gate sequence. If the residual $U(\delta)$ is too great, however, the algorithm proceeds to the next level. A further exhaustive search of the space of sequences of length up to $l_0$ is performed, this time in order to find the best approximation to $U(\delta)$. A subtlety at this step in the algorithm is that we do not have a closed form for $U(\delta)$, so need to find an approximation for $GS^\dagger(0)$ instead. The algorithm performs this by decomposing further into $VWV^\dagger W^\dagger = GS^\dagger(0)$ where $V$ and $W$ are the unitary gates that are then searched for. The sequence $VWV^\dagger W^\dagger$ that is closest to $U(\delta)$ is then returned by the search, and so the first-level approximation becomes
\begin{equation} S(1) = VWV^\dagger W^\dagger \ S(0)\label{length}\end{equation}

\noindent Note that $V,W,S(0)$ are all sequences of length up to $l_0$; the sequence $S(1)$ is therefore of length up to $5l_0$.

If $||G-S(1)||\le \epsilon$ then the algorithm terminates and returns $S(1)$. If not, the previous step is repeated to find a decomposition of the residual $U(\delta_1) = GS^\dagger(1)$. This is repeated until a sequence of the desired accuracy is found.

 As we can see from equation (\ref{length}), in the standard Solovay-Kitaev algorithm, the length of the approximating sequence grows by a factor of 5 at each level of recursion. As a result, the algorithm can only produce the approximating sequences of the length in the set $S_l=\lbrace l_0, 5l_0, 25l_0, 125l_0, ...\rbrace$, where $l_0$ is the length of the approximating sequence for the basic stage of the decomposition algorithm. Therefore, the vast majority of possible approximating sequences, which are not in $S_l$, cannot be generated by the algorithm. Furthermore, the best strategy for finding a short gate sequence is to make the initial approximation length $l_0$ as large as possible; however, this then means that at every step in the algorithm the length of the approximating sequence grows dramatically.  Because of this, the most efficient approximating sequences are likely to be missed, leading to much longer sequences than are necessary to reach the desired accuracy of approximation.

\section{Search space expansion}

The technique we will use is to expand the search space at the first level of the standard algorithm so that sequences that are longer than $l_0$ are also covered. We assume that $l_0$ is the longest possible sequence space that our computational resources can exhaustively search. By improving the accuracy of this initial approximation, $\epsilon_0$, we reduce the residual to be approximated at the next level of the algorithm. It is then intuitively reasonable that this will reduce the number of recursion levels implemented to find a sequence accurate to a given $\epsilon$. Formally, the residual error at recursion level $n$ is given by 

\begin{equation}\epsilon_n  =  \frac{1}{c^2 } \left(\epsilon_0 c^2 \right)^{\left(\frac{3}{2}\right)^n} \end{equation}

 (with $c$ as above)\cite{SK:2005}. We can therefore conclude that  the more we reduce $\epsilon_0$, the smaller will be $\epsilon_n$; and therefore the sooner the algorithm will find a sequence $\epsilon_n \le \epsilon$.

The initial stage of the algorithm gives us the first approximation gate sequence $S(0)$, equation (\ref{so}). The residual distance, $\epsilon_0$, from the exact gate $G$ is 
\begin{equation} ||G-S(0)|| \le \epsilon_0\end{equation}

We now start our space expansion technique. For simplicity, we partition the sequence $S(0)$ into two equal halves (in fact, the procedure can be performed by splitting the sequence into any number of parts, which may be unequal). Each of these subsequences $S^{(1)}(0), S^{(2)}(0)$ is of length $l_0/2$. For example, for a sequence $S(0) = HT^\dagger S^\dagger T$ we would have $S^{(1)}(0) = HT^\dagger \, \ S^{(2)}(0) = S^\dagger T$.

We now search once again over the space of sequences of length $l_0$ \footnote{In practise we will search over this maximum sequence space, although the general technique works for any subset of sequences $l_1 \le l_0$.} to find approximations $\{Z_i^{(1)}(0)\}$ and $\{Z_j^{(2)}(0)\}$ within $\bar{\epsilon_0}$ of $S^{(1)}(0)$ and $S^{(2)}(0)$:

\begin{eqnarray} \forall i ||S^{(1)}(0) - Z_{i}^{(1)}(0)||  & \le &  \bar{\epsilon_0}\nonumber\\
 \forall j ||S^{(2)}(0) - Z_{j}^{(2)}(0)||  & \le &  \bar{\epsilon_0}
 \label{s1s2}\end{eqnarray}

where we define the search regions by 
\begin{equation} \bar{\epsilon_0} = 0.5\epsilon_0 \end{equation}

\begin{figure}[t]
  \begin{center}
   \includegraphics[width=7cm]{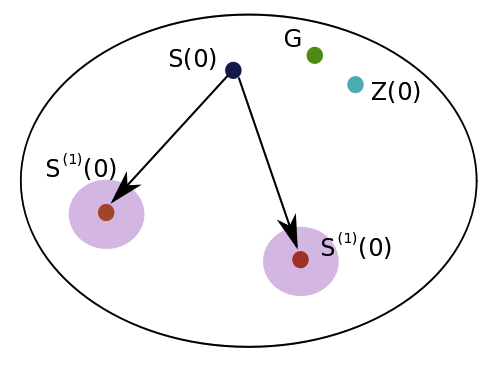}
  \end{center}
  \caption{Expanding the search space. Sequence $S(0)$ (a good first approximation to gate $G$) is split into two halves, $S(0) = S^{(1)}(0)S^{(2)}(0)$. Points in regions of distance $\bar{\epsilon}_0$ (shaded) away from $S^{(1)}(0), S^{(2)}(0)$ are recombined, and the closest to $G$, $Z(0)$ is chosen. With high probability, $||G-S(0)||>||G-Z(0)||$.}
  \label{search} 
\end{figure}

Figure \ref{search} shows this procedure schematically.

\IncMargin{1em}

\begin{algorithm*}
\caption{Space\_Expansion}
\SetKwInput{Input}{Input}
\SetKwInOut{Output}{Output}
\SetKwFunction{distance}{distance}
\SetKwFunction{SpaceExpansion}{Space\_Expansion}
\SetInd{0.5em}{1em}
\SetNlSkip{1em}
\Input{$G\in SU(2)$: Target of approximation}
\Input{$S$: Universal set of gates}
\Input{$S_0$: Stored set of sequences of gates from $S$ of length $l_0$}
\Input{$S_1$: Stored set of sequences of gates from $S$ of length $l_1$}
\Input{$\epsilon_0$: Accuracy of initial approximation}
\Input{$\bar{\epsilon_0}$: Accuracy of loop-internal approximation}
\Input{$k$: The desired cardinality of the set to be returned. $k=1$
  when Space\_Expansion is used as a standalone algorithm,
  $k>1$ when Space\_Expansion is used as a component of
  Recursive\_Space\_Expansion}

\Output{$k$ approximating sequences for $G$ of length $2l_1$}
\BlankLine
\SpaceExpansion($G,S,S_0,S_1,\epsilon_0,\bar{\epsilon_0},k$)\\
$R \leftarrow \lbrace r \in S_0 \vert \text{ }\distance(r,G) \leq \epsilon_0 \rbrace$\;
\ForEach{$r \in R$}{
  Split $r$ into 2 sub-sequences of the
  same length $\dfrac{l_0}{2}$, called $r_{\textrm{pre}}$ and $r_{\textrm{suf}}$\;
  $R_1 \leftarrow \lbrace r_1 \in S_1 \vert \text{ }\distance(r_1,r_{\textrm{pre}}) \leq \bar{\epsilon_0} \rbrace$\;
  $R_2 \leftarrow \lbrace r_2 \in S_1 \vert \text{ }\distance(r_2,r_{\textrm{suf}}) \leq \bar{\epsilon_0} \rbrace$\;
  Join $R_1$ and $R_2$ to have the following set: \\
  $R_3 \leftarrow \lbrace r_1r_2 \vert \text{ } r_1\in R_1\text{ and } r_2 \in R_2 \rbrace$\;
}
$R_4 \leftarrow \text{$k$ best approximations for G} \in R_3$\;
\Return{$R_4$}\\
\end{algorithm*}
\DecMargin{1em}

We now form the set of sequences 
\begin{equation} \{ Z_{ij}(0) = Z_i^{(1)}(0)Z_j^{(2)}(0)\} \end{equation}

These sequences are of length $2l_1$, and are derived from $S(0)$. We then search the sequences $Z_{ij}(0)$ to find the one, $Z(0)$, with minimal distance to the actual gate $G$, 
\begin{equation} ||G -Z(0)|| = \bar{\epsilon}_{\mathrm{min}}\end{equation}

 We now wish to show that, in general, 
 \begin{equation} \bar{\epsilon}_{\mathrm{min}} < \bar{\epsilon}(0)+\bar{\epsilon}(0) = \epsilon(0) \end{equation} 
 
 that is, that we have found a sequence $Z(0)$ of length $2l_1$ that is a closer approximation to $G$ than was the sequence $S(0)$ of length $l_0$, given by the original Solovay-Kitaev algorithm. To do this, consider any four gate sequences $U_{1,2}, \ V_{1,2}$. Given that 
 \begin{equation} ||U_1 - V_1||\leq \epsilon _1 \ \ \mathrm{and} \ \ ||U_2 -V_2||\leq \epsilon _2\end{equation}

  we have that 
 \begin{equation} ||U_1U_2 - V_1V_2||  \leq  ||U_1U_2 - V_1U_2|| + ||V_1U_2 - V_1V_2||, \end{equation}
 
 therefore
 \begin{eqnarray}
 {} & {} & ||U_1U_2 - V_1U_2|| + ||V_1U_2 - V_1V_2|| \nonumber\\
 & = & ||U_1 -V_1|| + ||U_2 - V_2|| \nonumber\\
  {} & \leq & \epsilon _1 + \epsilon _2.
\end{eqnarray}
  
We can therefore conclude that 
\begin{equation}
||U_1U_2 - V_1V_2|| \leq \epsilon _1 + \epsilon _2.
 \end{equation}

Since $\epsilon_1$ and $\epsilon_2$ are relatively small ($U1$ is a close approximation for $V_1$ and $U_2$ is a close approximation for $V_2$), it is possible to say that a sequence  $U_1U_2$ is a relatively close approximation for $V_1V_2$. If each $U \in SU(2)$ is presented by the specified vector $u \in \mathbb{R}^3$ or equivalently specified point in 3-D space (see below for further details of this mapping), $U_1U_2$ is in the sphere of centre $V_1V_2$ and radius $\epsilon_1+\epsilon_2$. With a large number of sequences $U_1$ and $U_2$ and their combinations  $U_1U_2$, the probability of having a very close approximation for $V_1V_2$ is relatively high. 

 We therefore conclude that 
 
 \begin{equation} \bar{\epsilon}_{\mathrm{min}} < \epsilon(0).\end{equation}
 
 for our new approximating sequence $Z(0)$ -- that is, we have strictly reduced the error in approximation by doubling the length of the approximating sequences, but not going on to the next level of recursion of the standard algorithm, which would increase it by a factor of 5.

This technique as it stands we call \emph{search space expansion} (SSE). At the initial stage of the SK algorithm the search space is expanded once, which significantly reduces the residual and hence the number of subsequent levels of recursion needed. Algorithmically, it can be written in the pseudocode form given in Algorithm 1.



\begin{algorithm*}
\caption{Recursive\_Space\_Expansion}
\SetKwInput{Input}{Input}
\SetKwInOut{Output}{Output}
\SetKwFunction{distance}{distance}
\SetKwFunction{SpaceExpansion}{Space\_Expansion}
\SetKwFunction{RecursiveSpaceExpansion}{Recursive\_Space\_Expansion}
\SetInd{0.5em}{1em}
\SetNlSkip{1em}
\Input{$G\in SU(2)$: Target of approximation}
\Input{$S$: Universal set of instructions}
\Input{$S_0$: Stored set of instruction sequences of gates from $S$ of length $l_0$}
\Input{$S_1$: Stored set of instruction sequences of gates from $S$ of length $l_1$}
\Input{$\epsilon_0$: Accuracy of initial approximation}
\Input{$\epsilon_1$: Parameter for Space\_Expansion function, indicates the accuracy of initial approximation for the called Space\_Expansion function}
\Input{$\bar{\epsilon_1}$: Parameter for Space\_Expansion function, indicates accuracy of loop-internal approximation for the called Space\_Expansion function}
\Input{$k$: Cardinality of the set of sequences that will be requested from Space\_Expansion $k>1$}
\Output{Approximating sequence for $G$ of length $4l_1$}
\BlankLine
\RecursiveSpaceExpansion($G,S,S_0,S_1,\epsilon_0,\bar{\epsilon_0},k$)\\
$R \leftarrow \lbrace r \in S_0 \vert \text{ }\distance(r,G) \leq \epsilon_0 \rbrace$\;
\ForEach{$r \in R$}{
  Split $r$ into 2 sub-sequences of the
  same length $\dfrac{l_0}{2}$, called $r_{\textrm{pre}}$ and $r_{\textrm{suf}}$\;
  $R_1 \leftarrow \lbrace r_1 \in \SpaceExpansion(r_{\textrm{pre}},S,S_0,S_1,\epsilon_1,\bar{\epsilon_1},k) \rbrace$\;
  $R_2 \leftarrow \lbrace r_2 \in \SpaceExpansion(r_{\textrm{suf}},S,S_0,S_1,\epsilon_1,\bar{\epsilon_1},k) \rbrace$\;
  Join $R_1$ and $R_2$ to have the following set: \\
  $R_3 \leftarrow \lbrace r_1r_2 \vert \text{ } r_1\in R_1\text{ and } r_2 \in R_2 \rbrace$\;
}
$r_3 \leftarrow$ The best approximation for $G$ in $R_3$\;
\Return{$r_3$}
\end{algorithm*}

We can, however, also apply SSE itself recursively at this initial stage, to get an even better $S(0)$ approximation. Rather than performing a standard search over sequences of length $l_1$ to find our approximations for $S^{(1)}(0)$ and $S^{(2)}(0)$ (equation \ref{s1s2}), we use SSE itself to find better approximations. So we use $S^{(1)}(0)$ and $S^{(2)}(0)$ as query gates for 2 SSE procedures, remaining the search spaces as in the standard SSE.

 The final approximation for $S(0)$ is therefore a sequence $ Z^\prime(0)$ of length $4l_1$, where
 
 \begin{equation} Z^\prime(0) = Z_a^{(1)}(0)Z_b^{(1)}(0)Z_a^{(2)}(0)Z_b^{(2)}(0) \label{zed}\end{equation}
 
  We call this technique \emph{recursive search space expansion} (recursive SSE), given in pseudocode form in Algorithm 2. We have now expanded the lookup space much further than was possible with the original Solovay-Kitaev algorithm, allowing for a much denser search of the space of possible approximating sequences.  The structure of the algorithm remains unchanged, and as a consequence, so too does the scaling of the accuracy of the approximation with the length of the sequences. The length $l$ of a sequence for accuracy $\epsilon$ is still given by
\begin{equation} l = O\left(log^c\left(1/ \epsilon\right)\right)\end{equation}
  
 \noindent with $c \approx 3.97$. However, this technique should significantly reduce the prefactor in scaling.

\section{Increasing lookup efficiency using geometric search}

The technique we have just described is very powerful in extending the set of searched sequences without requiring an exhaustive search over all possible approximations. However, there is an additional search cost for these methods that is not present in the Solovay-Kitaev algorithm as commonly used.

Firstly, each time SSE is invoked (either on its own or as part of a recursive SSE step), the space of sequences of length $l_0$ needs to be searched to find the regions of sequences that are distance $\bar{\epsilon}_0$ away from the subsequences $S^{(1,2)}(0)$. Each time SSE is used, an additional two searches are required to find the desired regions. Secondly, whenever sequences are combined to form a longer approximation to $G$, the list of combined sequences needs to be searched to find the one that is closest to $G$. This happens once per use of SSE or recursive SSE.

The first of these is by far the largest cost in our technique, as it is can occur many times in a given use of the decomposition algorithm. The second is only incurred once per level of recursion in the standard algorithm. We can keep the second cost tolerable by not increasing the number of times SSE is used recursively on itself -- for this reason, we describe recursive SSE as only splitting the sequence twice; any more, and the search cost to find the best $Z(0)$ (equation \ref{zed}) would be prohibitive. Without this restriction, we could use recursive SSE many times on itself to find sequences of arbitrary length, and have no need for the structure of the original Solovay-Kitaev algorithm. However, given the exponentially increasing cost of this search, we chose instead to restrict recursive SSE to two applications of SSE, and to then proceed to the next level of Solovay-Kitaev recursion if further accuracy in the decomposition is required.

In order for our techniques to be useful in feasible computational time, we need to find a way of performing the region-finding search in SSE efficiently. The most straightforward way is to search over the entire space of sequences up to length $l_0$ and pick out those within distance $\bar{\epsilon}_0$. Such a linear search is, however, very inefficient: in general, the search time will be exponential in $l_0$. However, we are searching a very structured space and should be able to make use of this structure in order to increase the efficiency of this search step. We will show now how to convert our matrix search problem into a 3D geometric search problem, and how we can then use the existing technique of geometric nearest-neighbour access trees (GNATs) to solve the search problem much more efficiently. We can use such a geometric technique in the original Dawson and Nielsen algorithm for the decomposition as well, and in the next section we will use it instead of a linear search when comparing our SSE techniques with the original algorithm. More sophisticated search techniques at this step are beginning to be developed, including the database searches of \cite{mssk} and \cite{mosc}. A further alternative use of geometric search techniques for the Dawson-Nielsen algorithm has been examined simultaneously, and came to our attention after our work was completed \cite{dombar}.


We can convert any matrix $SU(2)$ into a unique vector in a ball in a 3D real vector space of radius $2\pi$ and centred on the origin \cite[Ch.5]{introgt}. We can uniquely write any $U\in SU(2)$ as 
\begin{equation} U = \me^{\frac{-\mi}{2} \mathbf{v}\cdot \mathbf{\sigma}} \longrightarrow u(\mathbf{v}) \end{equation}

where $\mathbf{v} \in \R^3$ and $\mathbf{\sigma}$ is formed of the Pauli matrices,
\begin{equation} \mathbf{\sigma} = \oth{\sigma_x}{\sigma_y}{\sigma_z} \end{equation}

We therefore can fully and uniquely specify the matrix $U$ by specifying the vector $\mathbf{v}$. The distance function that we are using to search is now no longer the trace norm between two matrices $U_1, U_2$, but rather the norm between the two specifying vectors given by the Euclidean distance function.\\

We therefore have a 3D geometric search problem of finding vectors within the ball of radius $2\pi\bar{\epsilon}_0$ centred on the vector corresponding to the matrix $S^{(1,2)}(0)$. Such geometric searches in real space have been extensively studied, and we can therefore pick from the existing techniques the one that best suits our purposes.

\subsection{GNAT search}

\begin{figure}[t]
  \begin{center}
   \hspace*{20pt}
   \includegraphics[width=3.05in,height=2.4in]{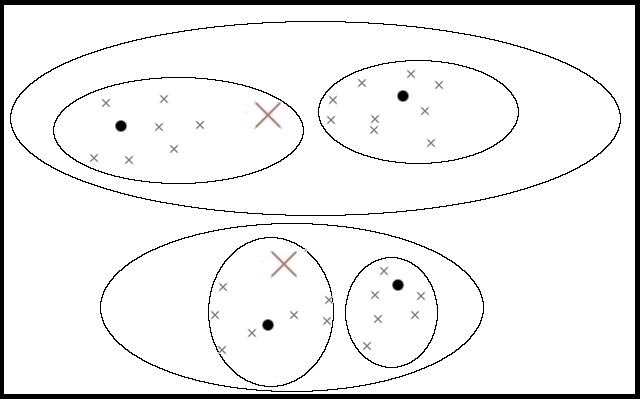}
  \end{center}
  \caption[A simple GNAT with clusters]{A simple GNAT with clusters. The points with surrounding circles are the splitting points of the same level clustering.}
  \label{gn:im} 
\end{figure}The standard method for increasing the efficiency of a real-space search is to use a \emph{tree-based} approach. By dividing the search space into clusters, tree-based searches are capable of reducing the computational complexity of a search over $n$ entries from $O(n)$ in the case of the straightforward linear search to $O(log(n))$ at least in the best or average case. The increasing efficiency comes from the ability of the search algorithms to skip entire clusters that are evidently not going to contain a correct answer. Amongst the most popular approaches are those of R-trees and $k$-d trees \cite{sellis1987r+,Bentley:1975:MBS:361002.361007}. These approaches divide the search space into clusters specified by coordinates in the space. These approaches were tried, and did not yield improved search results. Instead, we found that geometric nearest neighbour access trees (GNATs) are very effective. 

In a GNAT search, the space is partitioned into clusters around a number of fixed points called \emph{splitting points} \cite{GN:1995}. How best to choose these splitting points is an open research question; we use the common approach of picking them at random. At the initial stage of processing, each vector in the space is then analysed in turn to find its closest splitting point, and the distance from this point. The set of vectors that are closest to a given splitting point is the \emph{cluster} associated with that point. The cluster size can be defined by the greatest and least distance from all vectors in the cluster to the splitting point. Each cluster, if desired, can in turn be partitioned using splitting points to produce sub-clusters. This procedure can be applied recursively as many times as desired, until the size of each cluster is small enough that a linear search within the cluster is feasible.
Figure \ref{gn:im} shows clusters around splitting points in a 2D space.  
\begin{figure}[t]
  \begin{center}
   \includegraphics[width=3.5in,height=3in]{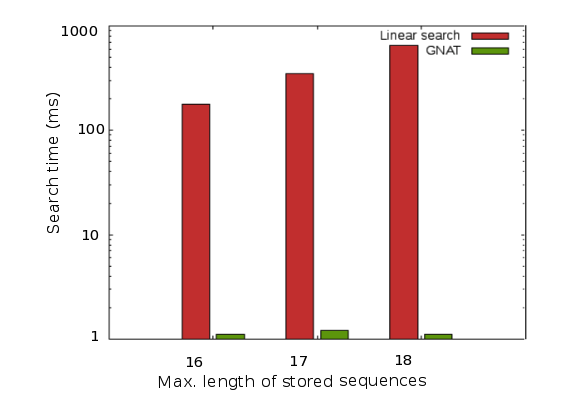}
  \end{center}
  \caption{Time required: GNAT vs. linear search, for stored sequence databases with $l_0 = 16,17,18$.}\label{gnat}
  \end{figure}

Calculating and storing the data associated with each point (which cluster it belongs to, and how far away it is from the splitting point) and each cluster (maximum and minimum distance between splitting point and all data points in the cluster) is the most computationally expensive step in the search procedure. However, when we use this procedure in our modified Solovay-Kitaev algorithm, we only need perform this step a single time for each set of library gates. This can be performed offline before the algorithm begins, and the structured data can be re-used for any decomposition problem using that library set.

The search proceeds as follows. Suppose we are searching for all the nearest neighbours $ \lbrace p \rbrace$ that are within distance $\epsilon$ of a given query point $x$, ie  $D(p,x) \leq \epsilon$ where $D$ is our distance function. We start by looking at the top-level clusters, with splitting points  $s_i$. The distance  between the splitting point and the query point is $D(s_i,x)=d_i$. By then applying the triangle inequality, we can see that the points $ \lbrace p \rbrace$ can only belong to clusters where the distance between the point $p$ and its splitting point satisfies
\begin{equation}d_i-\epsilon \leq D(p,s_i) \leq d_i+\epsilon \end{equation}

 All other clusters can be rejected, and the search at the next level concentrated on those that remain.

We can see how GNAT significantly improves the search time over a linear search. Figure \ref{gnat} shows the average time in ms to find the $\epsilon$-region around a query point for data sets with different volumes. These data are the average over 120 runs for each sized search space, choosing different query points each time. Note in particular that the time axis is on a log scale -- GNAT clearly outperforms linear search by two orders of magnitude.

  \begin{figure}[t]
\includegraphics[width=3.5in,height=2.8in]{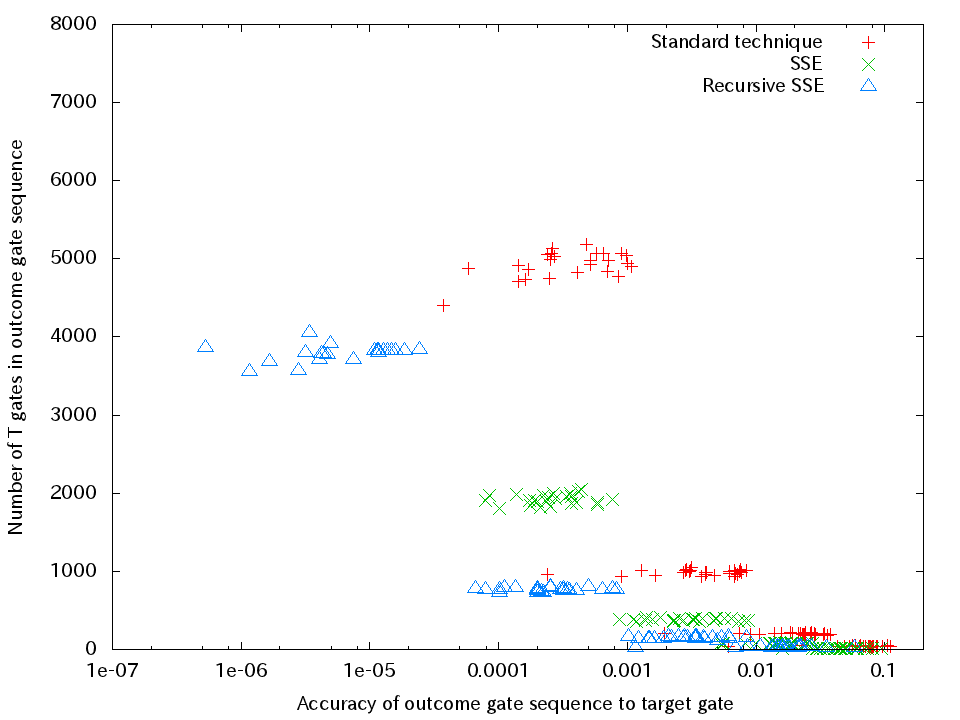}
\caption{Approximation accuracy vs. length of best approximating gate sequence found (number of $T$ gates)}
\label{pl:nor}
\end{figure}
  \begin{figure}[t]
\includegraphics[width=3.5in,height=2.8in]{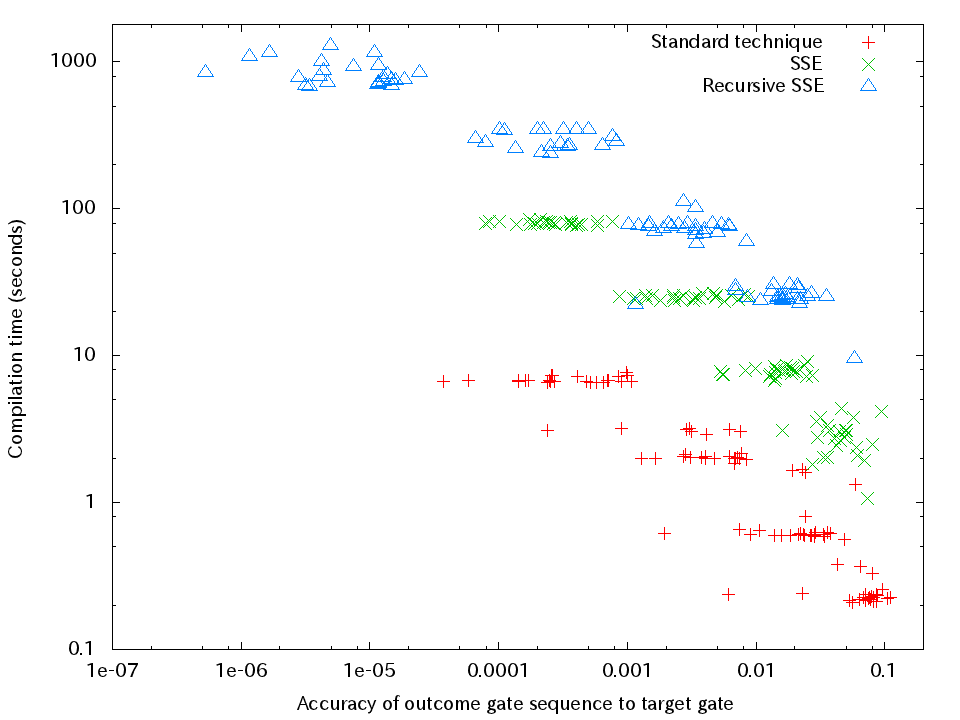}
\caption{Approximation accuracy vs. classical compilation time to find best approximating gate sequence}
\label{pl:not}
\end{figure}

\section{Comparison with the original algorithm}

In order to test the modifications to the Solovay-Kitaev algorithm, we implemented the original, original+SSE, and original+recursive SSE algorithms to find the length of gate sequences generated for a given gate and level of accuracy.  A set of  25 different randomly-generated matrices in $SU(2)$ was generated, and then approximations found for each matrix using the three different algorithms. The library gate set used was the minimal set $\{H,T,T^\dagger\}$, and the stored sequences were of length up to $l_0 = 18$. For each gate the best approximation at each of the first four levels of Solovay-Kitaev recursion was recorded. In all cases, GNAT search was used to find $\epsilon$-regions in the space of sequences of length up to $l_0$.

The largest quantum cost in fault-tolerantly implementing a sequence approximating the desired gate is, in general, in implementing the $T$ gates (as these require magic state injection). We therefore take as the appropriate measure of length for an approximating sequence the number of $T$ gates in the sequence. Figure  \ref{pl:nor} shows the length (number of $T$ gates) vs. the accuracy of the approximation for the best approximating sequence found by each algorithm, for each of the 25 randomly generated unitary matrices. The step-like behaviour in all three cases is caused by the recursion levels of the original algorithm. Both SSE and recursive SSE are significant improvements over the original algorithm, with recursive SSE clearly the better of the two. In both cases the length of sequences for a given approximation accuracy is reduced -- for example, when the accuracy required is around $10^{-4}$ then SSE alone reduces the length of the best sequence by a factor of three, and recursive SSE reduces it by a factor of seven. The number of levels of Solovay-Kitaev recursion also reduces significantly, from $n=4$ using the original algorithm to $n=3$ with SSE and to $n=2$ with recursive SSE.

Recursive SSE is therefore clearly better than the standard Solovay-Kitaev algorithm at producing gate sequences that cost less to implement in terms of quantum resources on a quantum computer. However, this is not the only consideration: we must also take into account the \emph{classical} processing time needed to find the best approximating sequence in each of the cases.  Figure \ref{pl:not} shows this classical pre-processing time vs. accuracy of the resulting approximating sequence. As the two plots are of the same dataset, they may be directly compared through their $x$-axes. As we would expect, the more computationally intensive recursive SSE algorithm takes a much longer time to run. However, note the log-scale on the time axis; the curve for recursive SSE is in fact sub-exponential in its scaling. 

We can therefore conclude that there is a straightforward trade-off between reducing the quantum cost of implementing a given single-qubit unitary, and the classical pre-processing time to find the sequence. In general, given the relative state of the two technologies, we would prioritize decreasing the quantum cost at the expense of classical processing. It is also important to bear in mind that sequence-finding can be run offline, before the algorithm starts, whereas implementing the approximating sequence in terms of quantum gates is by definition online. We therefore have an algorithm that gives us a shorter approximating sequence, using classical processing that scales less strongly than \cite{austindecomp}. This is therefore a middle-ground between such a flat, linear search, and the structured, sparse search of the standard recursive Solovay-Kitaev algorithm.

\section{Conclusion}
We have given a modified version of the Solovay-Kitaev algorithm that greatly reduces the length of the sequences used to approximate a unitary single-qubit gate. Our technique also reduces the number of levels of recursion required by the algorithm to reach a given level of accuracy. By reducing the depth of the quantum circuit used to approximate a given gate, we are then able to reduce the amount of error correction needed for fault-tolerant implementations of quantum algorithms. The cost for this is an increase in the classical processing needed to find these shorter and less costly quantum sequences, but the use of structured GNAT searches enables this to be performed in time that scales sub-exponentially (and can also be performed offline before the quantum algorithm starts). By increasing the space that is searched at the initial level of recursion in the original algorithm, we are able to use the powerful method of the recursive steps, but without leaving so much of the search space unexplored between levels of recursion. 

\section{Acknowledgements}

CH and RV acknowledge useful discussions with Krysta Svore and Alex Bocharov. CH appreciates valuable discussions with Dominic Berry and Barry Sanders. This research was supported by the Japan Society for the Promotion of Science through its Funding Program for World-Leading Innovative R\&D on Science and Technology (FIRST Program).

\bibliographystyle{unsrt}
\bibliography{SKE.bib}

\begin{thebibliography}{10}

\bibitem{PreskillFT}
J.~Preskill.
\newblock Fault-tolerant quantum computation.
\newblock In H-K. Lo, T.~P. Spiller, and S.~Popescu, editors, {\em Introduction
  to Quantum Computation and Information}. World Scientific Publishing, 1998.
\newblock arXiv:quant-ph/9712048.

\bibitem{preskillreliable}
J.~Preskill.
\newblock Reliable quantum computers.
\newblock {\em Proceedings of the Royal Society of London. Series A:
  Mathematical, Physical and Engineering Sciences}, 454(1969):385--410, 1998.

\bibitem{shor}
P.~W. Shor.
\newblock Scheme for reducing decoherence in quantum computer memory.
\newblock {\em Physical Review A}, 52:R2493--R2496, 2011.

\bibitem{shordiv}
D.~P. DiVincenzo and P.~W. Shor.
\newblock Fault-tolerant error correction with efficient quantum codes.
\newblock {\em Physical Review Letters}, 77:3260, 1996.

\bibitem{dummies}
S.~J. Devitt, Kae Nemoto, and W.~J. Munro.
\newblock Quantum error correction for beginners.
\newblock {\em arXiv:0905.2794v3 [quant-ph]}, 2011.

\bibitem{kitaev1}
A.~Yu. Kitaev.
\newblock Elementary gates for quantum computation.
\newblock {\em Russ. Math. Surv. 6}, 52:1191--1249, 1997.

\bibitem{kitaev2002classical}
A.~Yu. Kitaev, A.~Shen, and M.N Vyalyi.
\newblock {\em Classical and quantum computation, 1st edition}.
\newblock American Mathematical Society, 2002.

\bibitem{austindecomp}
A.~G. Fowler.
\newblock Constructing arbitrary {S}teane code single logical qubit
  fault-tolerant gates.
\newblock {\em Quantum Information and Computation}, 11:867--873, 2011.

\bibitem{GN:1995}
S.~Brin.
\newblock Near neighbor search in large metric spaces.
\newblock In {\em Proceedings of the 21th International Conference on Very
  Large Data Bases}, pages 574--584. Morgan Kaufmann Publishers Inc., 1995.

\bibitem{raussendorfprl}
R.~Raussendorf and J.~Harrington.
\newblock Fault-tolerant quantum computation with high threshold in two
  dimensions.
\newblock {\em Phys. Rev. Lett.}, 98:190504, 2007.

\bibitem{barenco}
A.~Barenco, C.~H. Bennett, R.~Cleve, D.~P. DiVincenzo, N.~Margolus, P.~Shor,
  T.~Sleator, J.~A. Smolin, and H.~Weinfurter.
\newblock Elementary gates for quantum computation.
\newblock {\em Phys. Rev. A}, 52:3457--3467, 1995.

\bibitem{harrow2002efficient}
A.~W. Harrow, B.~Recht, and I.L. Chuang.
\newblock Efficient discrete approximations of quantum gates.
\newblock {\em Journal of Mathematical Physics}, 43:4445, 2002.

\bibitem{SK:2005}
C.~M. Dawson and M.~A. Nielsen.
\newblock The {S}olovay-{K}itaev algorithm.
\newblock {\em Quantum Information and Computation}, 6(1):81--95, 2006.

\bibitem{mssk}
A.~Bocharov and K.~M. Svore.
\newblock A depth-optimal canonical form for single-qubit quantum circuits.
\newblock {\em arXiv:1206.3223 [quant-ph]}, 2012.

\bibitem{mosc}
M.~Mosca V.~Kliuchnikov, D.~Maslov.
\newblock Fast and efficient exact synthesis of single qubit unitaries
  generated by clifford and t gates.
\newblock {\em arXiv:1206.5236v2 [quant-ph]}, 2012.

\bibitem{dombar}
D.-S. Wang, M.~C. de~Olivereira, D.~W. Berry, and B.~C. Sanders.
\newblock Private communication.

\bibitem{introgt}
Y.~Kosmann-Schwarzbach.
\newblock {\em Groups and Symmetries: From Finite Groups to Lie Groups}.
\newblock Springer, 2010.

\bibitem{sellis1987r+}
T.K. Sellis, N.~Roussopoulos, and C.~Faloutsos.
\newblock The {R}+-tree: A dynamic index for multi-dimensional objects.
\newblock In {\em Proceedings of the 13th International Conference on Very
  Large Data Bases}, pages 507--518. Morgan Kaufmann Publishers Inc., 1987.

\bibitem{Bentley:1975:MBS:361002.361007}
Jon~Louis Bentley.
\newblock Multidimensional binary search trees used for associative searching.
\newblock {\em Commun. ACM}, 18(9):509--517, September 1975.

\end{thebibliography}

\end{document}